\documentclass[aps,amsmath,amssymb, preprint, 12pt]{revtex4-1}
\usepackage{amsmath, latexsym, color, graphicx, tabularx, booktabs, enumerate}
\usepackage{multirow}
\usepackage{dcolumn}
\usepackage{bm}
\usepackage{cleveref}
\usepackage{epstopdf}

\date{\today}

\usepackage[normalem]{ulem} 

\begin{document}
\author{Niraj K. Nepal$^{1*}$}
\author{Santosh Adhikari$^1$}
\author{Bimal Neupane$^1$}
\author{Shiqi Ruan$^1$}
\author{Santosh Neupane$^1$}
\author{Adrienn Ruzsinszky$^{1\dagger}$}
\affiliation{[1] Department of Physics, Temple University, Philadelphia, Pennsylvania 19122, United States}

\title{Understanding plasmon dispersion in nearly-free-electron metals: the relevance of exact constraints for novel exchange-correlation kernels within time-dependent density functional theory}
\begin{abstract}
 Small-wavevector excitations in Coulomb-interacting systems can be decomposed into the high-energy collective longitudinal plasmon and the low-energy single-electron excitations. At the critical wavevector and corresponding frequency where the plasmon branch merges with the single-electron excitation region, the collective energy of the plasmon dissipates into single electron-hole excitations. The jellium model provides a reasonable description of the electron-energy-loss spectrum (EELS) of metals close to the free-electron limit. The random phase approximation (RPA) is exact in the high-density limit but can capture the plasmonic dispersion reasonably even for densities with r$_s$ $>$ 1. RPA and all beyond-RPA methods investigated here, result in a wrong infinite plasmon lifetime for a wavevector smaller than the critical one where the plasmon dispersion curve runs into particle-hole excitations. Exchange-correlation kernel corrections to RPA modify the plasmon dispersion curve. There is however a large difference in the construction and form of the kernels investigated earlier. Our current work introduces recent model exchange-only and exchange-correlation kernels and discusses the relevance of some exact constraints in the construction of the kernel. We show that, because the plasmon dispersion samples a range of wavevectors smaller than the range sampled by the correlation energy, different kernels can make a strong difference for the correlation energy and a weak difference for the plasmon dispersion. This work completes our understanding about the plasmon dispersion in realistic metals, such as Cs, where a negative plasmon dispersion has been observed. We find only positive plasmon dispersion in jellium at the density for Cs. 
\end{abstract}
\maketitle

\section{Introduction}

\noindent Due to its computational feasibility and relatively high accuracy, approximate Kohn-Sham density-functional theory (KS-DFT) simulations are the basis of present-day first-principles computational materials science. Particle-particle interactions can require treatment beyond semilocal KS-DFT \cite{ORR02}. Experimental applications heavily rely on understanding and guiding electron-electron interaction. A relevant example of electron-electron interaction is scattering resulting in electron energy loss. Experimentally the electron loss spectrum can be realized by electron energy loss spectroscopy or inelastic scattering \cite{R06, SSSK95}.\\ 

Excited states can be accurately characterized by expensive Green’s function techniques. By virtue of the Runge-Gross theorem \cite{RG84}, DFT can be extended to time-dependent processes. Time-dependent density-functional theory (TDDFT) is becoming an attractive alternative to many-body perturbation theory and can offer, in principle, an unbiased and independent framework complementary to experimental observations, enabling the interpretation of specific experimental observations and predictions of new materials with targeted properties \cite{U11}.\\

Plasmon excitations are collective oscillations of electrons in the absence of an external electric field, that incorporate Coulomb interaction between electrons \cite{P64, GV05}. Due to the electron-electron interactions, plasmon excitations establish a high barrier when testing \textit{ab initio} theories. When the external perturbation is weak, linear response TDDFT \cite{PGG96} is a useful tool to describe optical excitation energies. In TDDFT the electron energy loss is quantified by the imaginary part of $ \chi $ , the spatially nonlocal and dynamic density-density response function. The poles of the interacting density-density response function contain information about the optical excitation energies. The same density-density response function can deliver further information about the plasmons for a range of wavevectors.\\  

Plasmon dispersion in nearly-free-electron alkali metals has attracted a great interest among experimentalists and theorists. The negative dispersion in the volume plasmon of the low-density alkali metals such as Rb and Cs has triggered a debate about the origin of the anomaly observed by most theoretical approximations within TDDFT and Fermi liquid theories \cite{VSF89}. A strong failure of approximations based on TDDFT with a static exchange-correlation kernel is the lack of a damping mechanism which results in an infinite lifetime of plasmons for a region of wavevectors smaller than the critical wavevector that separates plasmonic and particle-hole excitations \cite{VSF89, PN66}. The negative plasmon dispersion in low-density alkali metals can be attributed to
correlation effects or band structure \cite{AK94, KE99, T92, TS92}. There are pros and contras for both explanations in the literature \cite{T92, TS92}. In heavier alkali metals such as Cs, electron transition to the near-Fermi-level \textit{d} bands can occur \cite{AK94, KE99}. The transition energy of these electrons is comparable to the plasmon energy, potentially causing a negative dispersion. Additional corrections to the interacting response function or dielectric function can originate in a weak lattice potential and core polarization effects \cite{VSF89}.\\ 

The random phase approximation (RPA) is a Green’s function-based method that is often used to obtain the ground-state correlation energy of bulk and two-dimensional materials \cite{BP53, NRB18, NABR19, SGK13, HSK10}. Although RPA relies on the linear response TDDFT framework, its excitation energies are inaccurate because of the overestimated short-range exchange-correlation effects \cite{LP75}. The exact exchange-correlation kernel f$_{xc}$ that would provide these effects is a functional derivative of the exchange-correlation potential. RPA is often interpreted within DFT to have roots in the adiabatic connection fluctuation dissipation theorem \cite{LP77}. The bare RPA f$_{xc}$ $=$ 0 without a band structure does not yield a negative dispersion in heavy alkali metals \cite{AK94}. Theoretical predictions from the late 80's in polycrystalline metals showed negative dispersion for Cs beyond RPA when the band structure was included \cite{VSF89}. Some investigations of the role of correlation in the plasmon dispersion discuss kernels, but most tests consider the exchange-correlation effects at the level of adiabatic local density functional approximation (ALDA) only. The work of Tatarczyk et al. \cite{TSS01} steps beyond this limitation to some extent by considering some more model kernels based either on the uniform electron gas paradigm or on other constraints.\\

In our current work we aim to fill the gap in analyzing the plasmon dispersion with recent exchange-correlation kernels, beyond the early ones developed in the 90's. Our work aims to go beyond a simple analysis of exchange-correlation effects with kernels. The major goal here is to give an \textit{a priori }numerical analysis why kernels by themselves (without the band structure effects) can not predict negative dispersion in low density alkali metals. In this work, we rely on the jellium model. The dimensionless density parameters (r$_s$) for the jellium model corresponding to different metals are taken from Ref.~\cite{AM05}. We demonstrate that the exact constraints can lead to kernels that correctly predict a positive plasmon dispersion in jellium.

\section{Methodology: Exchange-correlation kernels within linear response TDDFT}

Nonempirical construction of density functionals has allowed widespread and successful applications of these approximations for the ground state \cite{SRZSRP16}. Various exact constraints such as the uniform electron gas limit, Lieb-Oxford bound or the one-electron limit are known for the ground state \cite{PRTSSC05}. According to linear response theory the interacting and noninteracting density-density response functions are coupled by the Dyson equation:

\begin{equation}
\chi _{ \lambda } \left( q, \omega  \right) =  \chi _{0} \left( q, \omega  \right) + \chi _{0} \left( q, \omega  \right)  \left(  \lambda v_{c} \left( q \right) +f_{xc}^{ \lambda } \left( q, \omega  \right)  \right)  \chi _{ \lambda } \left( q, \omega  \right),
\label{eq1}
\end{equation}

\noindent $\chi _{ \lambda } ( q, \omega )$  and   $\chi _{0} ( q, \omega )$  are the interacting and noninteracting response functions, respectively,  \( v_{c} \left( q \right) =\frac{4 \pi \lambda }{q^{2}} \)  and  \( f_{xc}^{ \lambda } \left( q, \omega  \right)  \)  are the Coulomb and exchange-correlation kernels. $ \lambda $  is the coupling constant that provides the adiabatic connection between a noninteracting Kohn-Sham ($\lambda$ $=$ 0) and the interacting real system ($\lambda$ $=$ 1) response. When Eq.~\ref{eq1} is applied to the uniform electron gas,  \(  \chi _{0} \left( q, \omega  \right)  \)  becomes the Lindhard function with complex frequencies \cite{L54}, a basic input to our current research. In the adiabatic approximation, the exact kernel is a second functional derivative of the ground state exchange-correlation energy. In practice the exact kernel is unknown but can be modeled by satisfying exact physical constraints. Kernels are related to the $``$local field factors$"$  as
G (q, $\omega$) $=$ $\frac{f_{xc}(q, \omega)}{-v_c (q)}$.\\

Many real systems have densities close to the paradigm uniform electron gas, as in alkali metals. The uniform electron gas is therefore a simple model system with physical relevance. All exchange-correlation kernels in this work model the uniform electron gas with known limiting behavior at q \(  \rightarrow 0 \)  and q \(  \rightarrow \infty. \)  The simplest approximation is known as the adiabatic local density approximation (ALDA) kernel for $\lambda$ $=$ 1 \cite{ZS80}:\\

\begin{equation}
f_{xc}^{ALDA} ( q \rightarrow 0,  \omega =0 ) =-\frac{4 \pi A}{k_{F}^{2}}
\end{equation}

\noindent with  \( A=\frac{1}{4}-\frac{k_{F}^{2}}{4 \pi }\frac{d^{2} \left( n \varepsilon _{c} \right) }{dn^{2}} \), where  \( k_{F}= \left( 3 \pi ^{2}n \right) ^{1/3} \)  is the Fermi wavevector and  \(  \varepsilon _{c} \)  the correlation energy per particle of the uniform electron gas.  \( A=\frac{1}{4} \)  belongs to the exchange-only ALDA, while the density-dependent term in  \( A \)  gives the correlation beyond the high-density limit.\\

The real-space representation of ALDA is a delta function which indicates the spatial locality of this kernel. ALDA gives reasonable accuracy for low-frequency, long-wavelength excitations, but is not the right choice for a general correction to RPA \cite{ZS80}. The ALDA kernel was applied to the ground state correlation energy of the uniform electron gas but makes an error of $ \sim $ 0.5 eV \cite{LGP00}. This error is the same in magnitude but of opposite sign to the error that RPA makes for the same system.\\

The ALDA kernel can be made nonlocal, by applying a cutoff that makes the exchange-correlation kernel cancel the Hartree kernel for q $>$ k\textsubscript{cut}. The cutoff is introduced by the renormalized ALDA (rALDA) expression \cite{OT12}

\begin{equation}
 f_{xc}^{rALDA} ( n,q) =- [ \theta ( k_{cut}-q ) \frac{4 \pi }{k_{cut}^{2}}+ \theta  ( q-k_{cut} ) \frac{4 \pi }{q^{2}} ]
\end{equation}

\noindent with the cutoff wavevector   \( k_{cut}=\frac{k_{F}}{\sqrt{A}}. \) \ \ By construction the rALDAxc kernel keeps the correct  q \(  \rightarrow 0 \)  limit of ALDA, but improves the wrong q \(  \rightarrow \infty \) \  behavior of ALDAxc. For inhomogeneous systems, more ingredients like the density gradient or the kinetic energy density give more flexibility for kernels, as for ground state density functional approximations. The kinetic energy density is one of the ingredients of the nonlocal energy optimized (NEO) exchange-only kernel \cite{BLR16}.\\

The NEO kernel improves the ground state correlation energy and structural properties of real systems beyond RPA \cite{BLR16}. The NEO kernel is designed to satisfy further physical constraints beyond both ALDA or rALDA, and has the form

\begin{equation}
f_{x}^{NEO}=-\frac{4 \pi }{2q^{2}} [ 1-e^{- \beta q^{2}/k_{F}^{2}} ]
\end{equation}

where  \(  \beta =\frac{1}{4\widetilde{c} \left( 1-z^{2} \right) }. \)   The one-electron limit is reached when the ingredient  \( z=\frac{ \tau^{w}}{ \tau} \) equals 1, where  \(  \tau^{w} \)  is the one-electron kinetic energy density \cite{W35} and  \(  \tau \)  the positive kinetic energy density constructed from the Kohn-Sham orbitals. In the uniform electron gas,  \( z \)  is zero. When q \(  \rightarrow 0, \) \ \  the NEO kernel is properly independent of q for the uniform electron gas. The parameter  \( \widetilde{c} \)  has a key relevance to the current work. In the construction of NEO,  \( \widetilde{c} \)  is designed to give a correction to the RPA correlation energy in the high-density limit. In other words, the standard  \( \widetilde{c}=0.264 \) \ parameter in NEO provides a unique fit to the exact second-order correlation energy for the spin-unpolarized electron gas.  The $``$second-order exchange$"$  contribution to the second-order correlation energy of the uniform gas is the correction to direct RPA from wavefunction anti-symmetry that survives in the high-density limit. It can be evaluated from explicit expressions given by von Barth and Hedin for RPA \cite{VH72} and by Langreth and Perdew \cite{LP75}  beyond RPA.\\

 \noindent \(  \beta  \)  can be chosen to satisfy another constraint relevant to the long-wavelength ( \( q \rightarrow 0 )  \)  limit. This choice is also made in the ALDA, rALDA, and in the CP07 dynamic exchange-correlation kernel constructed by Constantin and Pitarke \cite{CP07}. The compressibility sum rule, \cite{I82}, as

\begin{equation}
f_{xc} ( n;q \rightarrow 0,  \omega =0 ) =\frac{d^{2}}{dn^{2}} [ n \varepsilon _{xc} ( n )  ] ,
\end{equation}

\noindent is an important requirement for frequency-dependent exchange-correlation kernels. Satisfying the compressibility sum rule,  \(  \beta  \)  becomes

\begin{equation}
 \beta =\frac{1}{4\widetilde{\text{c}}}   =-\frac{2k_{F}^2}{4 \pi }\frac{d^{2}}{dn^{2}} [ n \varepsilon _{xc} ( n )  ] = 2A
\end{equation}
or   \( \widetilde{c}=0.5 \)  in the high-density limit.

Thus the energy optimized value of  \( \widetilde{c}=0.264 \)  in the high-density limit is different from the value  \( \widetilde{c}=0.5 \) \  that yields the correct small- \( q \)  kernel in the high-density limit. In the next section we will extensively discuss the impact of these physical constraints on the plasmon dispersion and provide a novel insight about the role of correlation effects. It will turn out that the difference between the ALDA and NEO kernels is important for the correlation energy, but not very important for the plasmon dispersion.\\

\begin{figure}[h!]
	\includegraphics[scale=0.4]{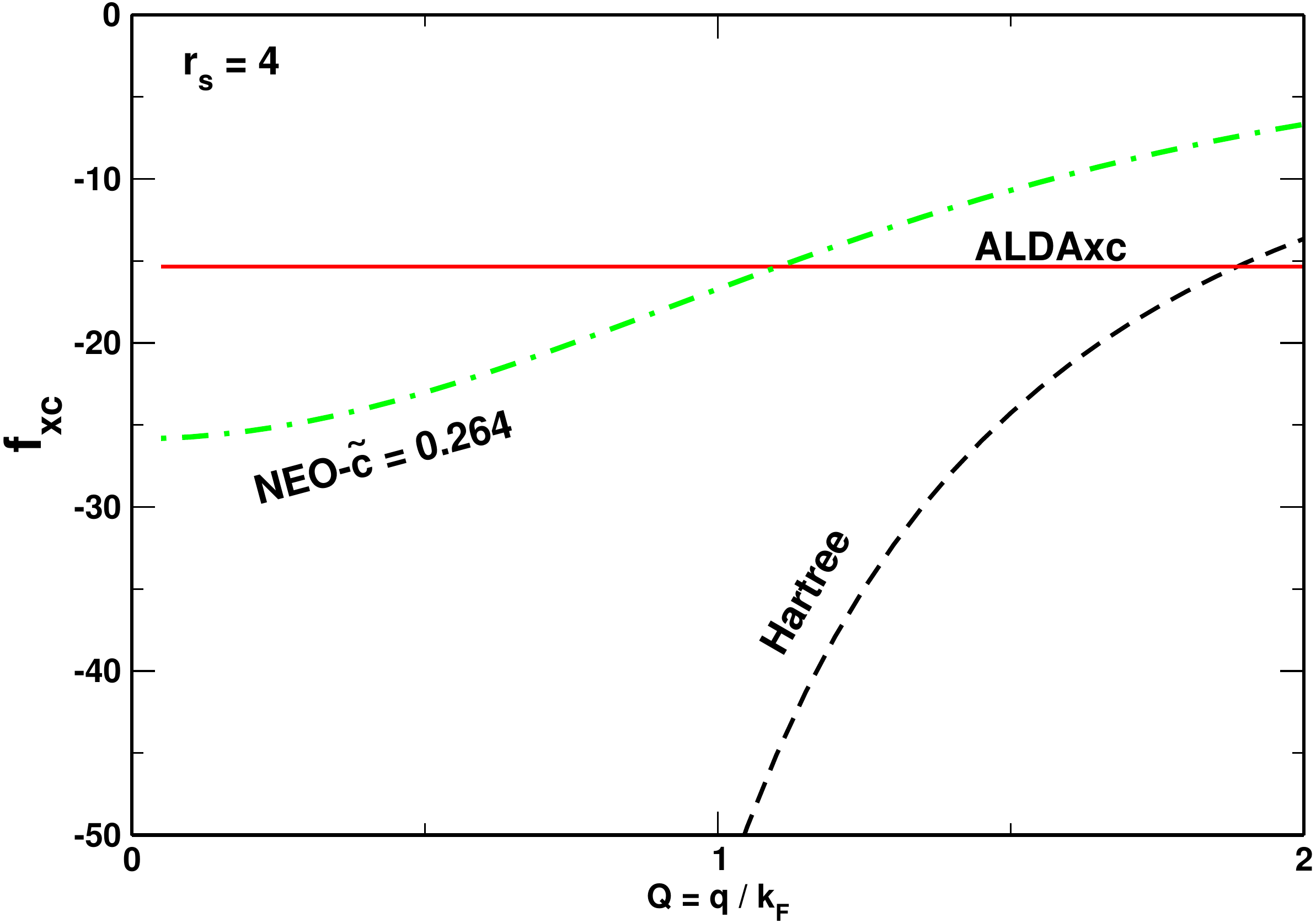}
	\caption{The Hartree kernel, ALDAxc kernel, and standard NEO xc kernel with $\widetilde{c}$ $=$ 0.264 vs Q $=$ q/k$_F$ for the uniform electron gas at r$_s$ $=$ $(\frac{3}{4\pi n})^{1/3}$ $=$ 4. For ease of comparison, it is actually minus the Hartree kernel (-4$\pi/\text{q}^2$) that is plotted here.}
	\label{fig0}
\end{figure}

To set the scale of the problem, we plot in Fig.~\ref{fig0} three kernels as functions of wavevector q: the ALDAxc kernel (exact at small q), the NEO xc kernel with $\widetilde{c}$ $=$ 0.264 (which we will argue later is more correct than ALDAxc at larger q), and minus the Hartree kernel (-4$\pi/\text{q}^2$). Clearly, in the wavevector region 0 $<$ q/k$_F$ $<$ 1 that shapes the plasmon dispersion, the Hartree kernel (the only one present in RPA) dominates over the xc kernel. This dominant effect of the Hartree kernel explains the overall good performance of RPA for plasmon dispersion. For wavevectors q/k$_\text{F}$ $>$ 1, which are important for the correlation energy, the xc kernels have a larger effect. An even better xc kernel might interpolate between ALDAxc at small q and NEO at larger q.

\section{Plasmon dispersion with spatially nonlocal exchange-correlation kernels in nearly-free-electron metals}

The alkali metals Na and K are nearly-free-electron (NFE) systems and therefore realizations of the uniform electron gas. Correlation effects in alkali metals can be strong enough, especially for low electron densities, to impact electronic excitations. The high-resolution electron energy loss experiments indicate that the plasmon dispersion in heavy alkali metals such as Rb and Cs becomes negative, i.e., the plasmon frequency decreases with increasing wavevector q \cite{VSF89}. Since all the above experiments and calculations were performed for periodic crystals, it is difficult to decouple correlation and band structure effects \cite{STLS68} in the dispersion of the plasmon excitations.\\

Further calculations by Ku and Eguiluz in K crystal \cite{KE99} indicate the relevance of band structure versus correlation and demonstrate that the exchange-correlation effects beyond RPA at the ALDA level have only a minor role in the dispersion of plasmons. A similar observation was made by Quong and Eguiluz  \cite{QE93} for Na crystal. Although\ crystal\ periodicity is important when modeling realistic conditions, the jellium model can offer an important way to separate the impact of many-body correlations and band structure. Within our work, we want to provide an additional theoretical and numerical insight only about the many-body correlation effects and explain why in general no beyond-RPA approximation for jellium can predict the correct plasmon dispersion and lifetime for Cs. The justification of our results is based only on exact physical constraints imposed on the construction of beyond-RPA approximations.  By using the jellium model for alkali metals, we can build upon the conclusion that ALDA gives a minor improvement beyond RPA for the plasmon dispersion. With the nonlocal exchange-correlation kernels developed since  the later 90's, we can make further conclusions about how these recent approximations compare to ALDA and RPA in terms of correlation effects.\\

We can make two groups of assessed approximations. The first group includes exchange-only and exchange-correlation kernels based on the ALDA approximations. ALDAx and ALDAxc are both local kernels but differ in correlation contribution. rALDAx and rALDAxc are nonlocal. The second group consists of NEO exchange-only kernels \cite{BLR16} with the $\widetilde{c}$ parameter constructed by satisfying different physical constraints. The default version of the NEO kernel yields the exact correction to the RPA correlation energy of jellium in the high-density limit:

\begin{equation}
 e_{c}^{2X}=\frac{3}{8 \pi ^{3}} \int _{0}^{\infty}dKK^{2}\widetilde{G}_{x} ( K )  \int _{0}^{\infty}dW \{ 2 b (K, W)  \} ^{2},
\end{equation}

\noindent where  \( \widetilde{G}_{x} \left( K \right)  \)  refers to the kernel, according to the correspondence between kernels and local-field factors.  \( K=\frac{q}{2k_{F}} \)  is a dimensionless wavevector, and  \( W=\frac{ \Omega }{2k_{F}^{2}} \) \ is a dimensionless frequency. The explicit expression for the second-order exchange energy   \( e_{c}^{2X} \) comes from Langreth and Perdew \cite{LP77} beyond RPA and uses the RPA correlation energy for the uniform electron gas given by von Barth and Hedin \cite{VH72}. The $\widetilde{c}$ parameter that corresponds to the second-order correlation energy was found to be 0.264.\\

\noindent Alternatively,\ in the long-wavelength limit we can use the compressibility sum rule formulated as   \( \frac{d^{2}}{dn^{2}} \left[ n \varepsilon _{xc} \left( n \right)  \right]  \)  to determine $``$ \( \widetilde{c} \) $"$ . Then the $\widetilde{\text{c}}$ parameter of NEO can be estimated from the compressibility sum rule of the ALDAxc expression. This fitting delivers a different  \( \widetilde{c} =  \) (0.43 - 0.47) at metallic densities considered here. While formally the NEO approximation remains an exchange-only kernel, this kind of fitting brings long-wavelength exchange-correlation effects into our NEO kernel \cite{CP07}. \\

Clearly $``$ \( \widetilde{c} \) $"$  controls the correlation or screening within the NEO approximation starting from  \(  $\( \widetilde{c} \)$ \rightarrow \infty \) \ in RPA. The impact of $``$ \( \widetilde{c} \) $"$  as a screening parameter on ground state correlation energies was established by Bates et al in 2016 \cite{BLR16}.  Changing $``$ \( \widetilde{c} \) $"$  from its default 0.264 value was shown to yield different correlation energies in the uniform electron gas at r\textsubscript{s} = 4. \\

The exchange-only NEO kernel can be explicitly turned into an exchange-correlation approach by replacing $``$ \( \widetilde{c} \) $"$  by an electron-density-dependent parameter. This approach was tested for jellium slab correlation energies at moderate densities, and resulted in improved integrated correlation energy \cite{RCP16}. For a given density, the density dependence can make $``$ \( \widetilde{c} \) $"$  significantly smaller than its default value. In our analysis we also investigate the effect of a low $``$ \( \widetilde{c} \) $"$  parameter on the plasmon dispersion of various NFE metals. For testing purposes we choose  \( \widetilde{c}=0.0037. \)  Notice that this choice of  \( \widetilde{c} \)  represents an unphysically low density according to Eq. (13) of Ref.~\onlinecite{RCP16}. \\

At first, we discuss the plasmon dispersion up to the wavevector region where plasmons decay into single-particle excitations. We consider all the exchange-correlation kernels described above. Within the static approximation for the kernel  \( f_{xc} \left( q \right) , \)  the plasmon frequency  \(  \omega _{p} \left( q \right)  \)  is found by solving the equation  \(  \epsilon  \left( q, \omega  \right) =1- \left( v_{c} \left( q \right) +f_{xc} \left( q \right)  \right)  \chi _{0} \left( q , \omega \right) =0, \)  for  \(  \lambda =1  \) \cite{TSS01}, and the solutions are undamped outside the particle-hole contiunuum, i.e., for  \( q<q_{c} \)  where  \( \frac{ \omega _{p} \left( q_c \right) }{ \varepsilon _{F}}=2 \left( \frac{q_{c}}{k_{F}} \right) + \left( \frac{q_{c}}{k_{F}} \right) ^{2}. \)  For the r\textsubscript{s }values considered here,  \( q_{c}/k_{F} \approx 1 \ll k_{cut}/k_{F} \approx 2, \)  where  \( k_{cut} \) \  is the cutoff wavevector for a kernel. Thus rALDA and ALDA kernels will yield the same plasmon dispersion.\\

Al is a metal with rather high density, and RPA becomes relatively exact in the high-density limit \cite{QE93}. Here, we model Al by jellium with r$_s$ $=$ 2.07. The small wavevector behavior is demonstrated by the correct plasmon energy known from an EELS experiment \cite{PO75}. All ALDA and rALDA kernels return the correct long-wavelength limit of f$_{xc}$. The $\widetilde{\text{c}}$ \(  =  \) 0.47\  NEO exchange kernel keeps the correct long-wavelength of f$_{xc}$. Since the compressibility sum rule delivers direct information about the long-wavelength limit, among all the approximations NEO $\widetilde{\text{c}}$ \(  =  \) 0.47 has a direct impact on the curvature of the dispersion. The fitting against this exact constraint designates that the plasmon dispersion must start out as horizontal at small wavevectors. NEO $\widetilde{\text{c}}$ \(  =  \) 0.47 exemplifies the best behavior in the long wavelength limit (q $\rightarrow$ 0) that any static kernel for jellium at r$_s$ $=$ 2.07 can demonstrate.\\

\begin{figure}[h!]	
	\includegraphics[scale=0.3]{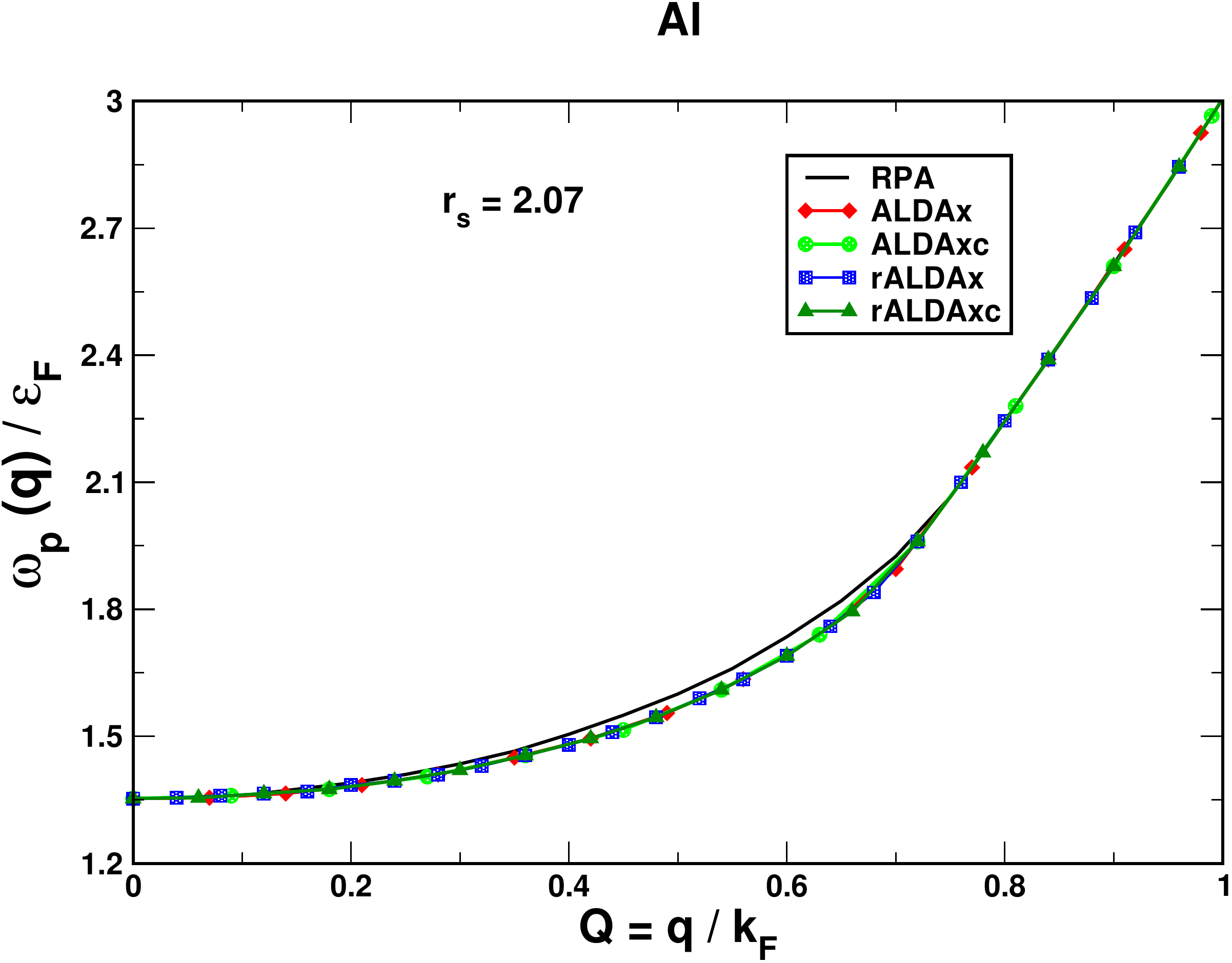}
	\includegraphics[scale=0.3]{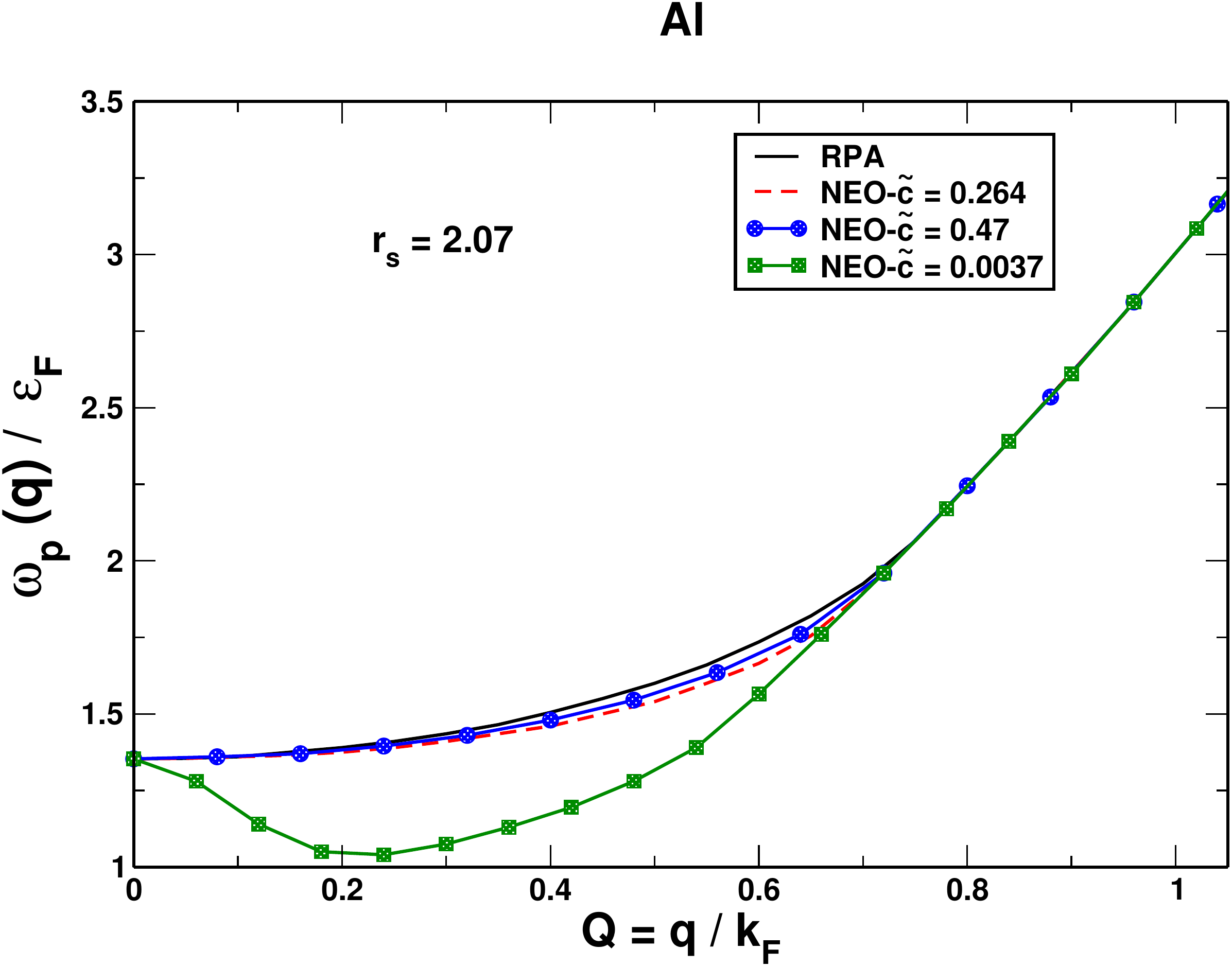}
	\caption{The plasmon dispersion for Al (modeled as jellium with r$_s$ $=$ 2.07) up to the critical wavevector. The left panel shows the dispersion obtained with RPA and beyond-RPA with ALDAx, ALDAxc, rALDAx and rALDAxc approximations. The right panel shows the dispersion from RPA and the three NEO approximations with the  \( \widetilde{c} \)  parameters corresponding to different choices.}
	\label{fig1}
\end{figure}

Comparing the ALDA and rALDA kernels in the left panel of Fig.~\ref{fig1}, it is apparent that the nonlocal feature of the rALDA relevant in the ground state correlation energy does not change the plasmon dispersion, and all ALDA and rALDA approximations yield the same dispersion curve. The NEO  \( \widetilde{c} \) = 0.264 and  \( \widetilde{c} \) = 0.47 methods basically agree, while the NEO \( \widetilde{c} \) = 0.0037 is completely unphysical with low plasmon frequencies as an indication of the lack of exact constraints (See Fig.~\ref{fig1}).


\begin{figure}[h!]	
	\includegraphics[scale=0.3]{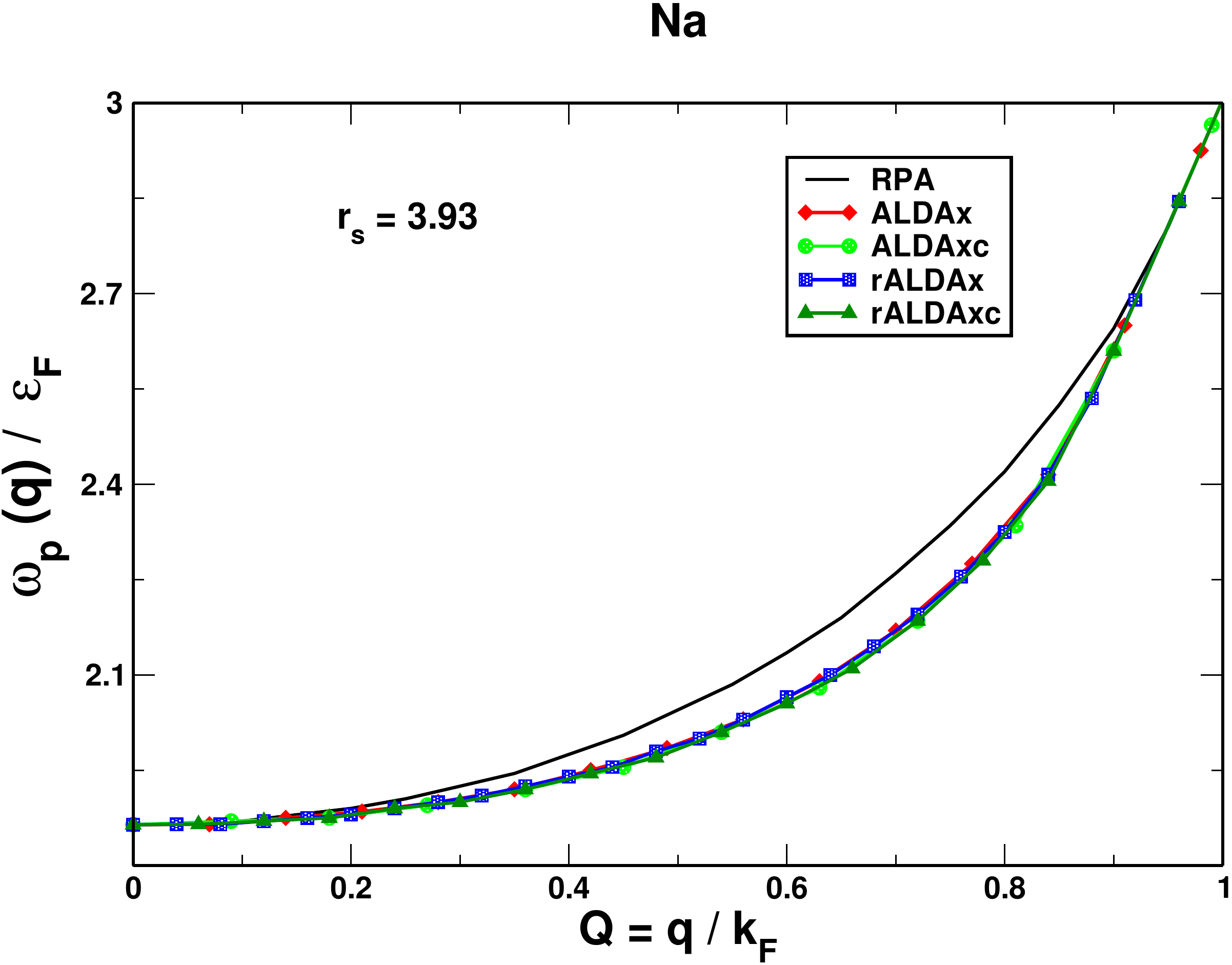}
	\includegraphics[scale=0.3]{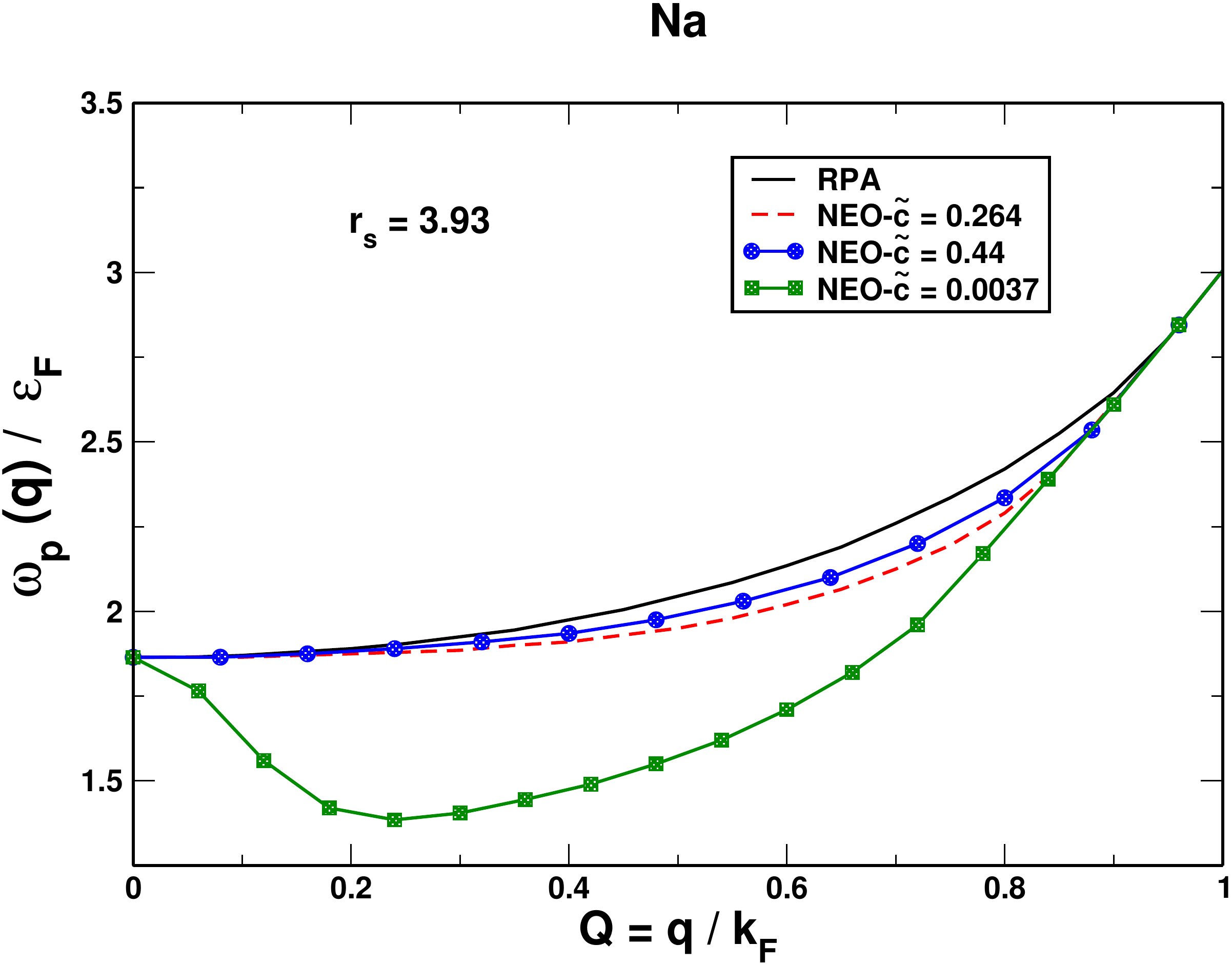}
	\caption{The plasmon dispersion for Na (modeled as jellium with r$_s$ $=$ 3.93), up to the critical wavevector. The left panel shows the dispersion obtained with RPA and beyond-RPA with ALDAx, ALDAxc, rALDAx and rALDAxc approximations. The right panel shows the dispersion from RPA and the three NEO approximations with the  \( \widetilde{c} \)  parameters corresponding to different choices.}
	\label{fig2}
\end{figure}

For jellium at r\textsubscript{s}=3.93 (as for Na), all ALDA and rALDA dispersion curves barely differ in the left panel of Fig.~\ref{fig2}. As in Al, there is no significant change coming from the nonlocal kernels. According to the right subfigure, the NEO  \( \widetilde{c}=0.264 \)  and NEO  \( \widetilde{c}=0.44 \)  kernels differ more than they do in Al for a range of $ \sim $ 0.5 for dimensionless wavevector Q. NEO  \( \widetilde{c}=0.0037 \)  leads to unphysically low plasmon frequencies. \\


\begin{figure}[h!]	
	\includegraphics[scale=0.3]{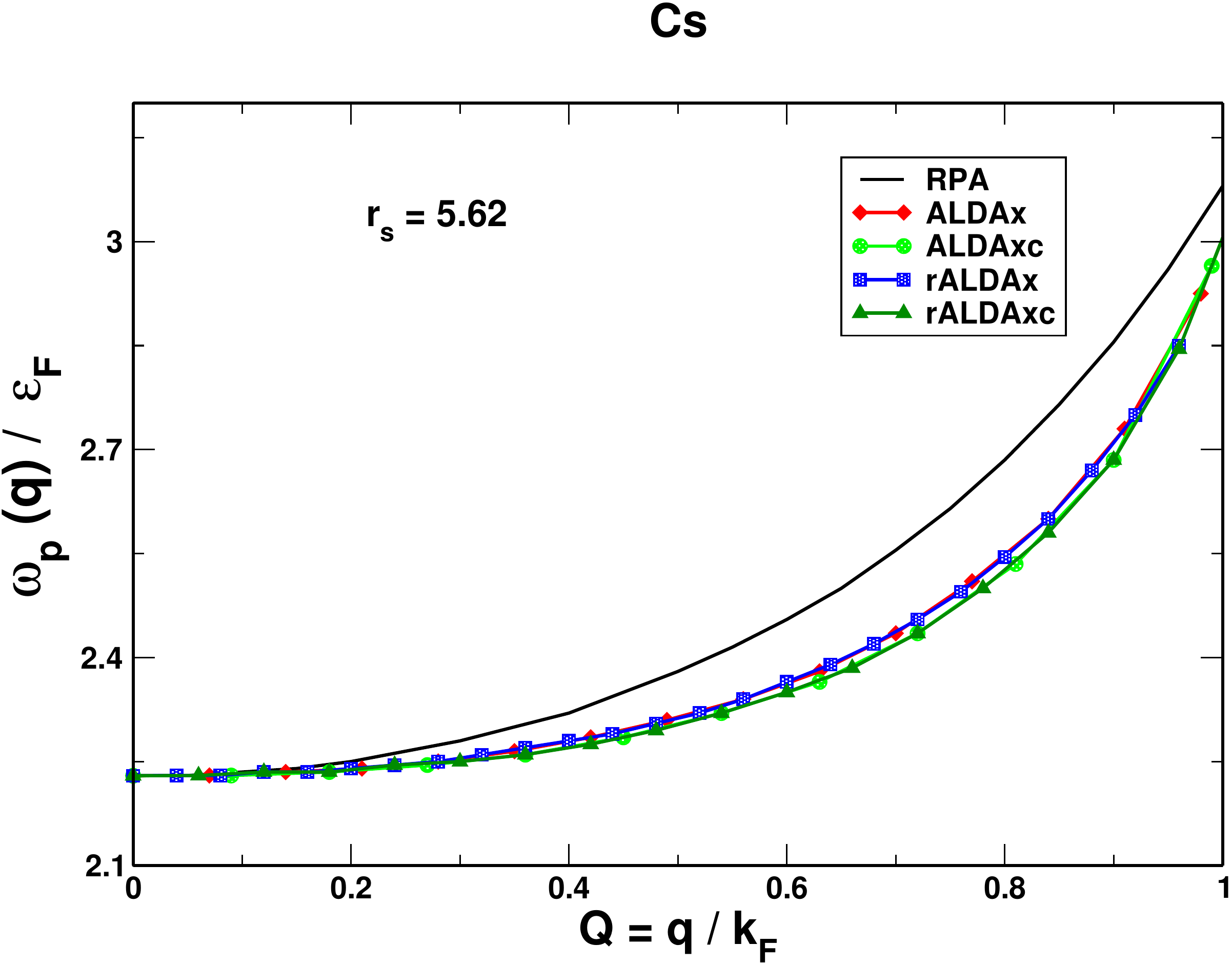}
	\includegraphics[scale=0.3]{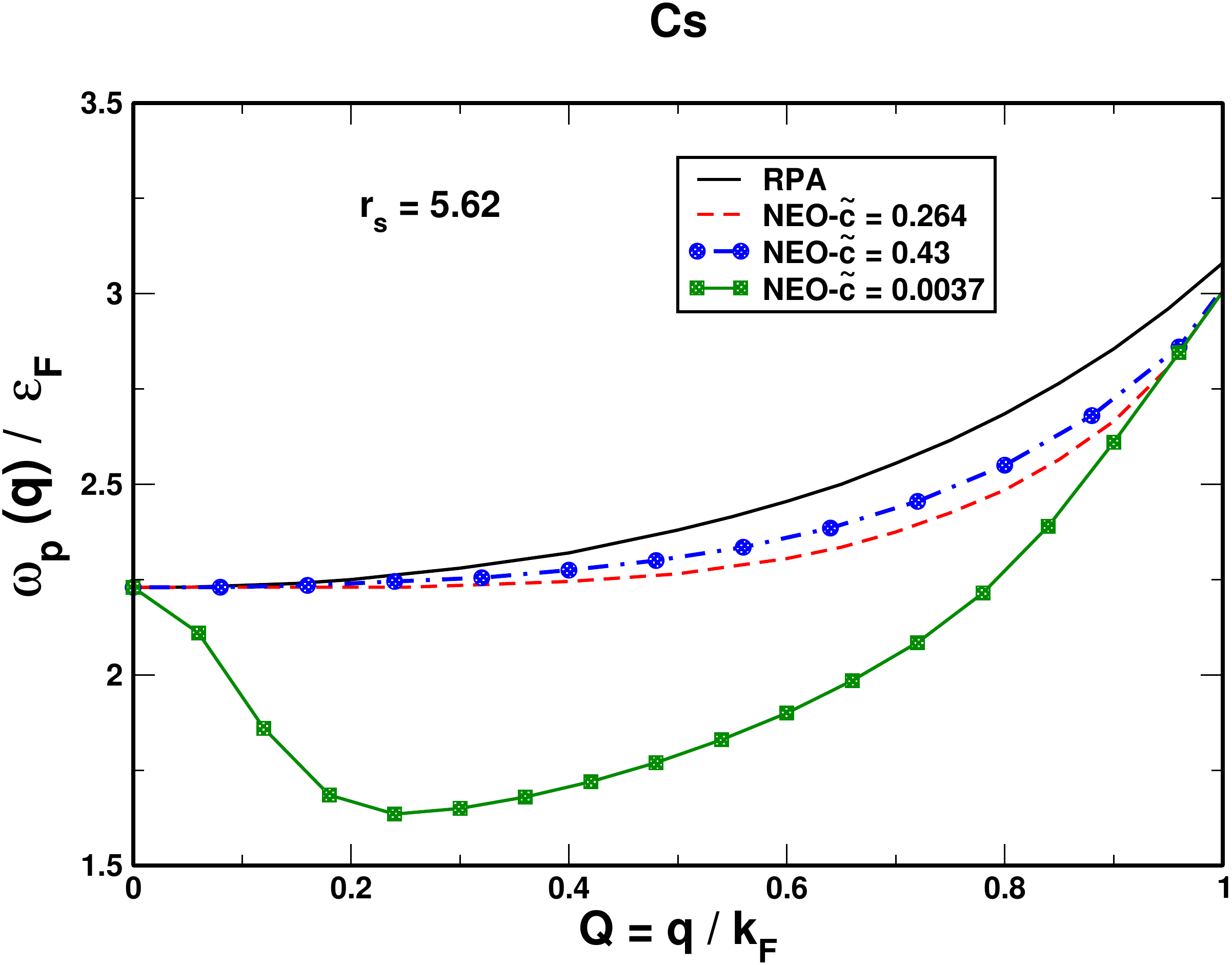}
	\caption{The plasmon dispersion for Cs (modeled as jellium with r$_s$ $=$ 5.62) up to the critical wavevector. The left panel shows the dispersion obtained with RPA and beyond-RPA with ALDAx, ALDAxc, rALDAx and rALDAxc approximations. The right panel shows the dispersion from RPA and the three NEO approximations with the $\widetilde{c}$ parameters corresponding to different choices.}
	\label{fig3}
\end{figure}

Cs is the alkali metal with the lowest density \cite{FSE97}. We modeled it here as jellium with r$_s$ $=$ 5.62. This characteristic manifests itself in the plasmon dispersion when comparing the approximations in the left and right panels of Fig.~\ref{fig3}. Being correct at small  \( q, \) \ the ALDAxc and rALDAxc  are more suitable than ALDAx for lower densities in Cs, but the nonlocality versus locality in rALDAxc and ALDAxc does not much affect the dispersion. Comparing the NEO approximations, NEO  \( \widetilde{c}=0.264 \)  results in more correction beyond-RPA than it does in the previous two metals. Furthermore NEO  \( \widetilde{c}=0.264 \)  yields more correction in the plasmon frequencies than any of the ALDA and rALDA kernels. The exact compressibility-sum-rule-based NEO  \( \widetilde{c}=0.43 \)  behaves more like ALDAx or ALDAxc. NEO  \( \widetilde{c}=0.0037 \)  considerably lowers the plasmon dispersion. At the first glance this could be mistaken for the observed behavior for the real low-density Cs.


\begin{figure}[h!]	
	\centering
	\includegraphics[scale=0.4]{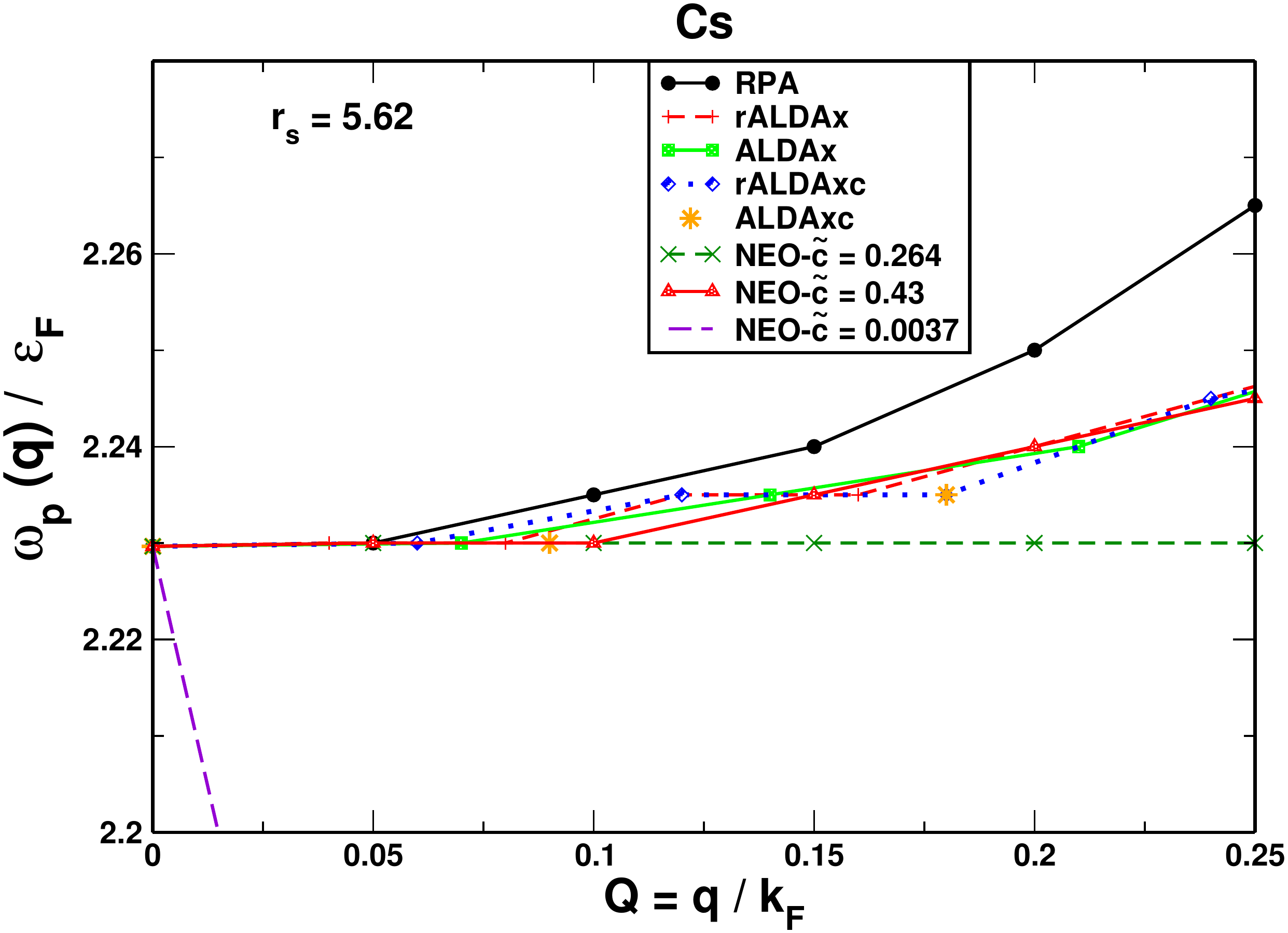}
	\caption{The long wavelength behavior or all approximations considered in this work.}
	\label{fig4}
\end{figure}

To summarize the role of the exact constraints, we compare all the exchange and exchange-correlation models described above. Figure ~\ref{fig4} displays the small wavevector behavior of all kernels and of RPA. The exact dispersion relation is known to be quadratic in the wavevector q as E $ \sim $   \( q^{2} \) . Except NEO  \( \widetilde{c}=0.0037, \)  all exchange and exchange-correlation kernels and RPA are properly horizontal at small Q wavevectors and become properly quadratic as Q increases. The horizontal line is a consequence of the exact physical constraints satisfied by these methods. NEO  \( \widetilde{c}=0.0037 \)  is not consistent with any exact constraint. While NEO  \( \widetilde{c}=0.264 \) \ is based on the exact high-density limit of the correlation energy, the unphysical NEO deviates from this constraint and picks up a wrong negative quadratic dispersion. \\

\begin{figure}[h!]
\centering
		\includegraphics[scale=0.32]{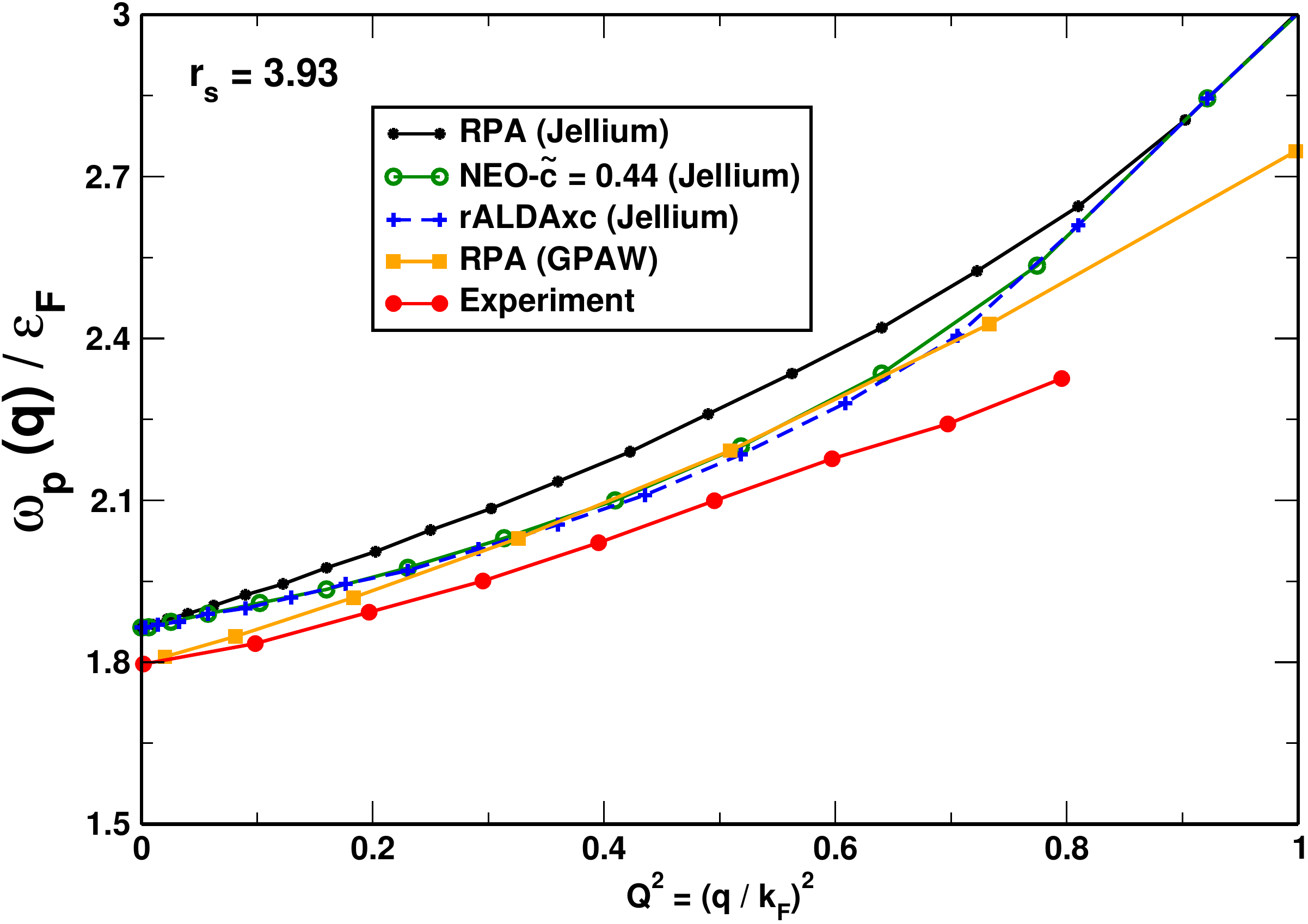}
		\includegraphics[scale=0.32]{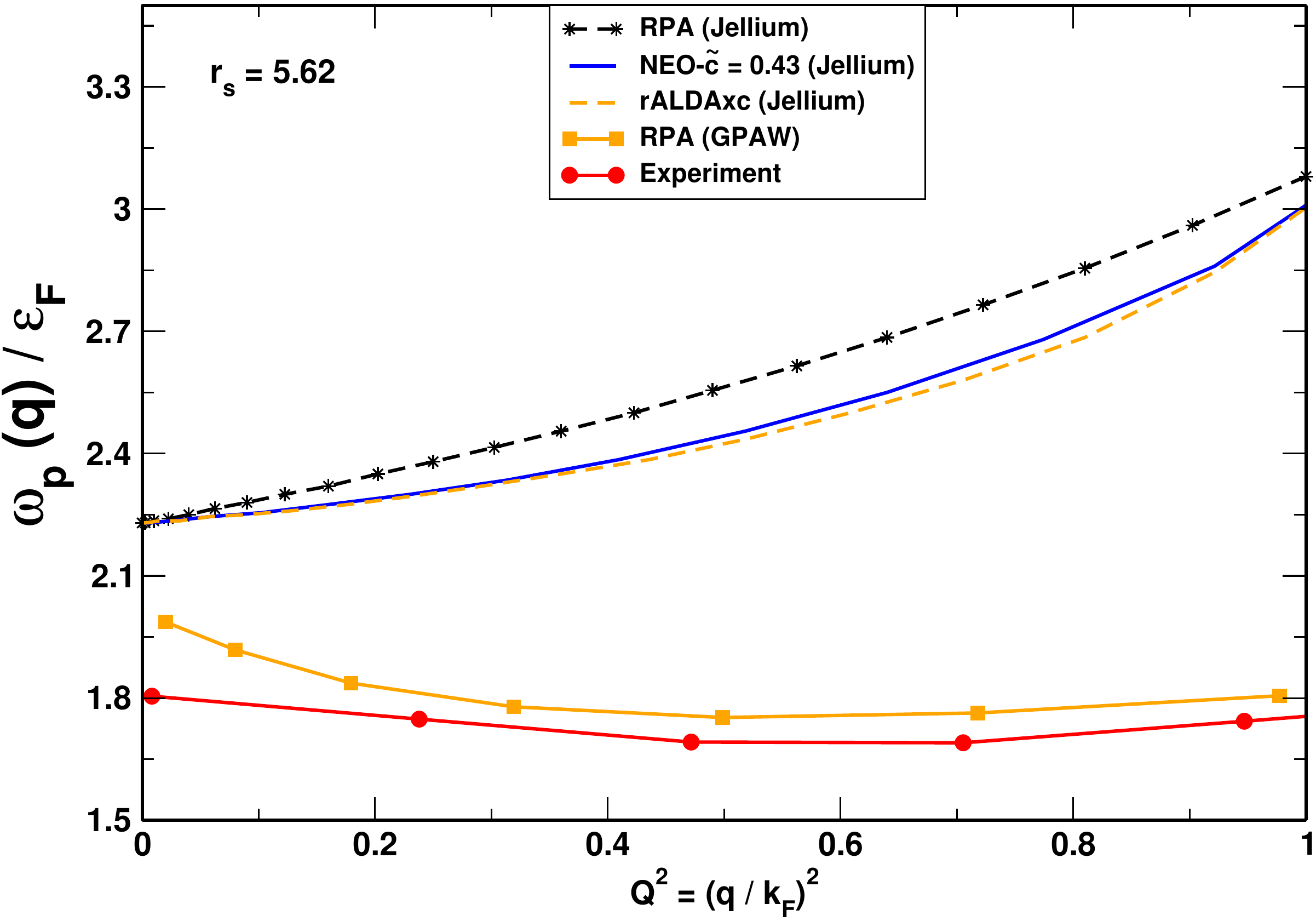}
\caption{The plasmon dispersion of Na (left) and Cs (right) with RPA and some exchange-correlation kernels within the jellium model. For both Na and Cs, the dispersion with RPA is also displayed with band structure effects obtained from the GPAW code, showing results close to experiment \cite{VSF89}. The momentum transfer (\textbf{q}) is in the [100] direction for bulk solids.}
\label{real}
\end{figure}

We have also extracted the plasmon dispersion for the real metals Na and Cs by calculating the electron energy loss spectrum using the GPAW\cite{gpaw1, gpaw2, gpaw_response, ase} code. The results are presented in Figure~\ref{real}. We used the projector-augmented wave (PAW)\cite{B94} pseudopotential provided with the GPAW code, an energy cutoff of 600 eV, and a 16$\times$16$\times$16 k-point mesh to sample the Brillouin zone. Our calculations including band structure effects confirm that the band structure significantly alters the plasmon dispersion for Cs, while its effect is negligible for Na. The plasmon frequency at q$\rightarrow$0 shows a ~0.5 eV renormalization compared to the RPA value within the jellium model. The effect of the band structure in Cs is significantly large enough to dominate over the changes from one kernel to another.

\section{The dynamic structure factor within and beyond-RPA}

The dynamic structure factor \cite{LP77} or spectral function S(q, $\omega$) shows the distribution of frequencies $\omega$ for density fluctuations of wavevector q in the ground state of the uniform electron gas. Although it arises from all density fluctuations of a given wavevector, the spectral function for small wavevector typically peaks around the frequency of a plasmon. The inverse frequency width of this peak, by the uncertainty principle, provides a lower bound on the decay time for such a fluctuation, while the dependence of the peak on q reflects the plasmon dispersion. Here we will investigate the effects of various model exchange-correlation kernels on the spectral function. The dynamic structure factor  \( S \left( q, \omega  \right)  \)  is proportional to  Im\(\chi\)  \cite{JP85}, the loss component of the dynamic density-density response function:
\begin{equation}
S (q, \omega) = -\frac{1}{\pi} Im\chi(q, \omega) \Theta({\omega})
\end{equation}
This quantity has been investigated by Lewis and Berkelbach \cite{LB19} using an equation-of-motion coupled cluster singles and doubles formalism, which unlike our TDDFT methods allows for plasmon decay via multi-pair electron-hole decay channels. With our real static exchange-correlation kernels, the plasmon at small finite q does not decay.\\

\begin{figure}[h!]	

	\includegraphics[scale=0.33]{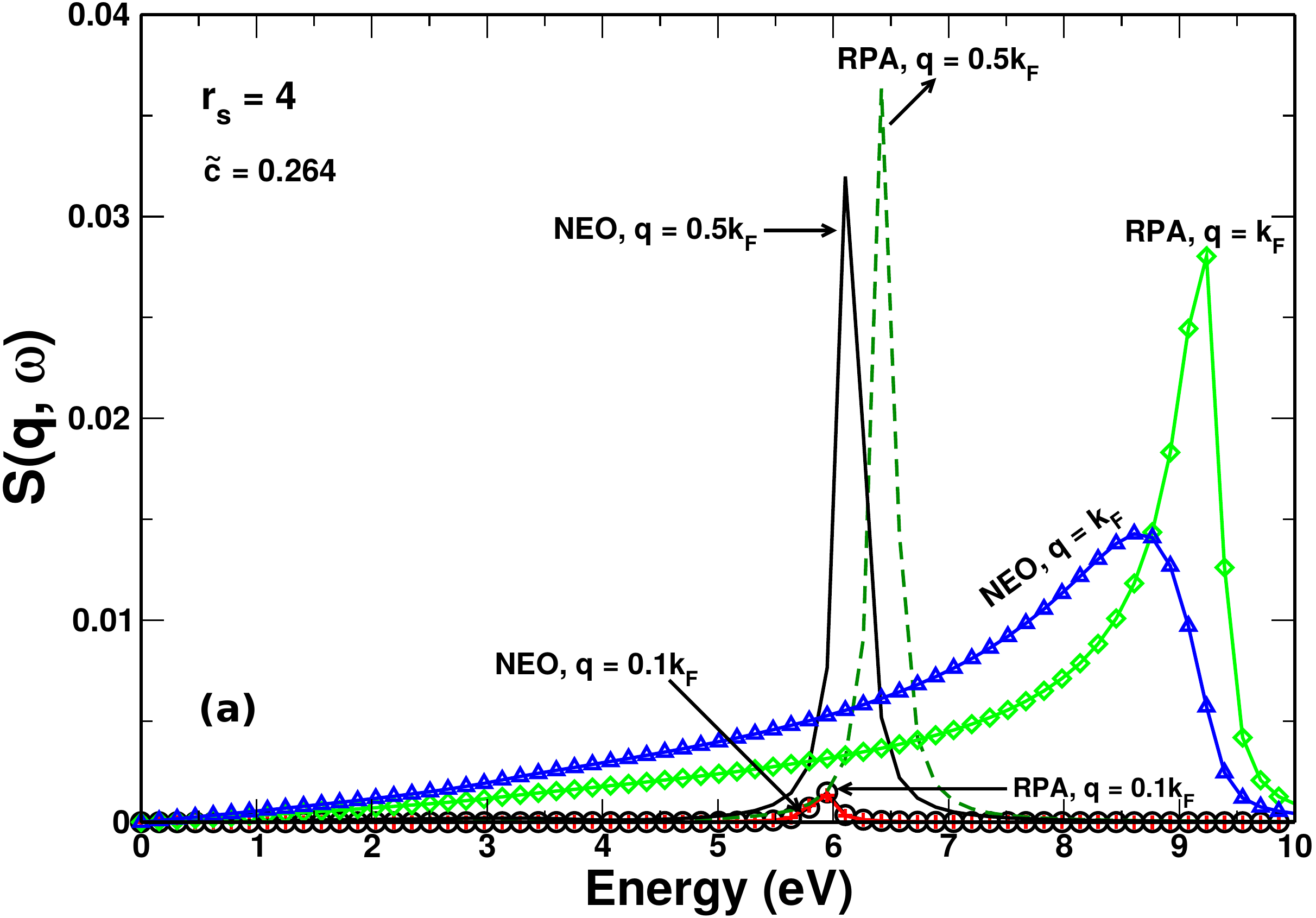}
		\includegraphics[scale=0.33]{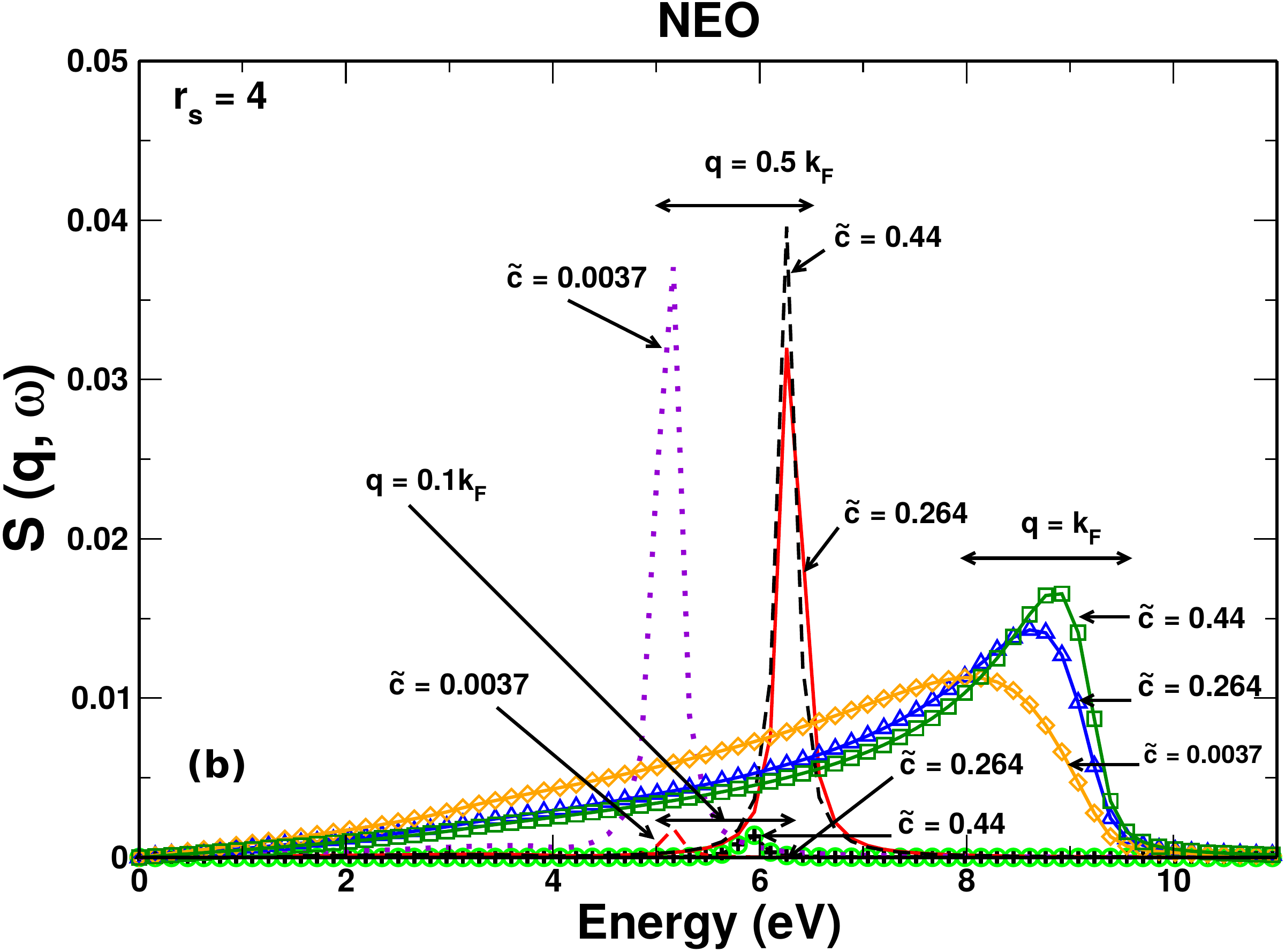}
		\vspace{1cm}
		\includegraphics[scale=0.35]{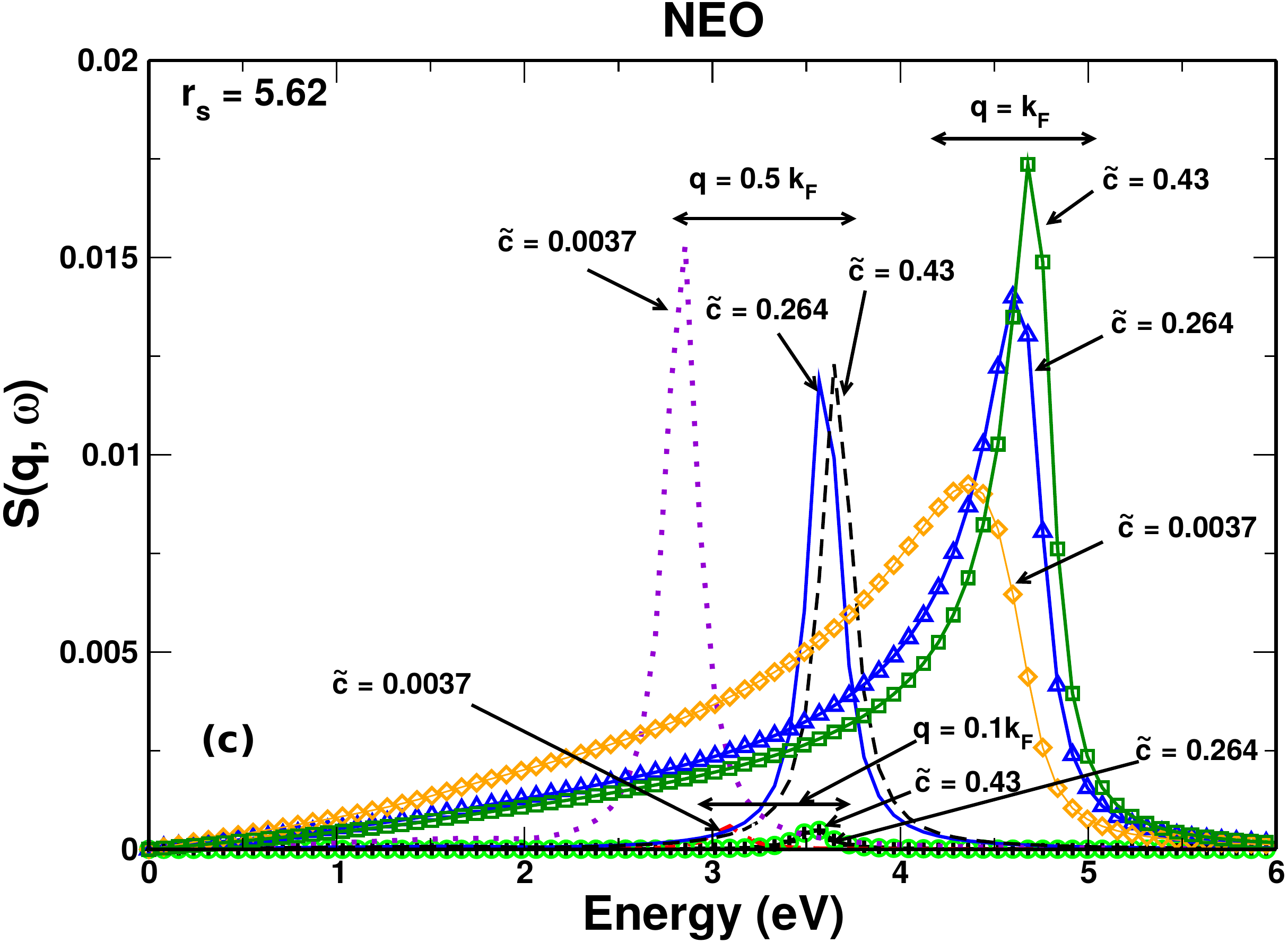}
	\caption{a: The jellium spectral functions for RPA and NEO\textbf{  \( \widetilde{c}=0.264 \) } at r\textsubscript{s }= 4. b: the spectral functions for three NEO kernels at r\textsubscript{s}=4. c: the jellium spectral functions for three NEO kernels at r\textsubscript{s}=5.62 corresponding to Cs.}
	\label{fig6}
\end{figure}

In Figure ~\ref{fig6}, we analyze our approximations at three wavevectors:  \( q=0.1k_{F}, \)   \( q=0.5k_{F}, \) \  and  \( q=k_{F}. \)  The latter wavevector is close to the one at which the plasmon decays into the single-pair electron-hole continuum at r\textsubscript{s} = 4. The first Figure~\ref{fig6} (a) compares RPA to the default NEO kernel with  \( \widetilde{c}=0.264. \)  The line shapes at  \( q=0.1k_{F} \)  and  \( q=0.5k_{F} \)  are very similar for both methods. At  \( q=k_{F} \)  the line shape from NEO becomes broader than the one from RPA, and the plasmon peak from NEO is shifted to a lower frequency. Figure ~\ref{fig6}.b compares three NEO kernels at r\textsubscript{s} = 4 for the same three wavevectors.\\


The static structure factor S(q) is the integral over frequency of S(q, $\omega$), divided by the electron number N. S(q) determines the well-known correlation energy of the uniform electron gas. We will now show that the range of q/k$_F$ (less than or about equal to 4) that distributes to the correlation energy is much greater than the range of q/k$_F$ (less than or about equal to 1) that contributes to the plasmon dispersion. In the smaller range (but not in the larger one), the ALDAxc kernel is nearly sufficient, while the default NEO kernel is nearly sufficient over the larger range. This completes our plasmon dispersion analysis with an explanation why a negative dispersion cannot exist in jellium at the density of Cs.\\

\begin{figure}[h!]	
	
	\includegraphics[scale=0.3]{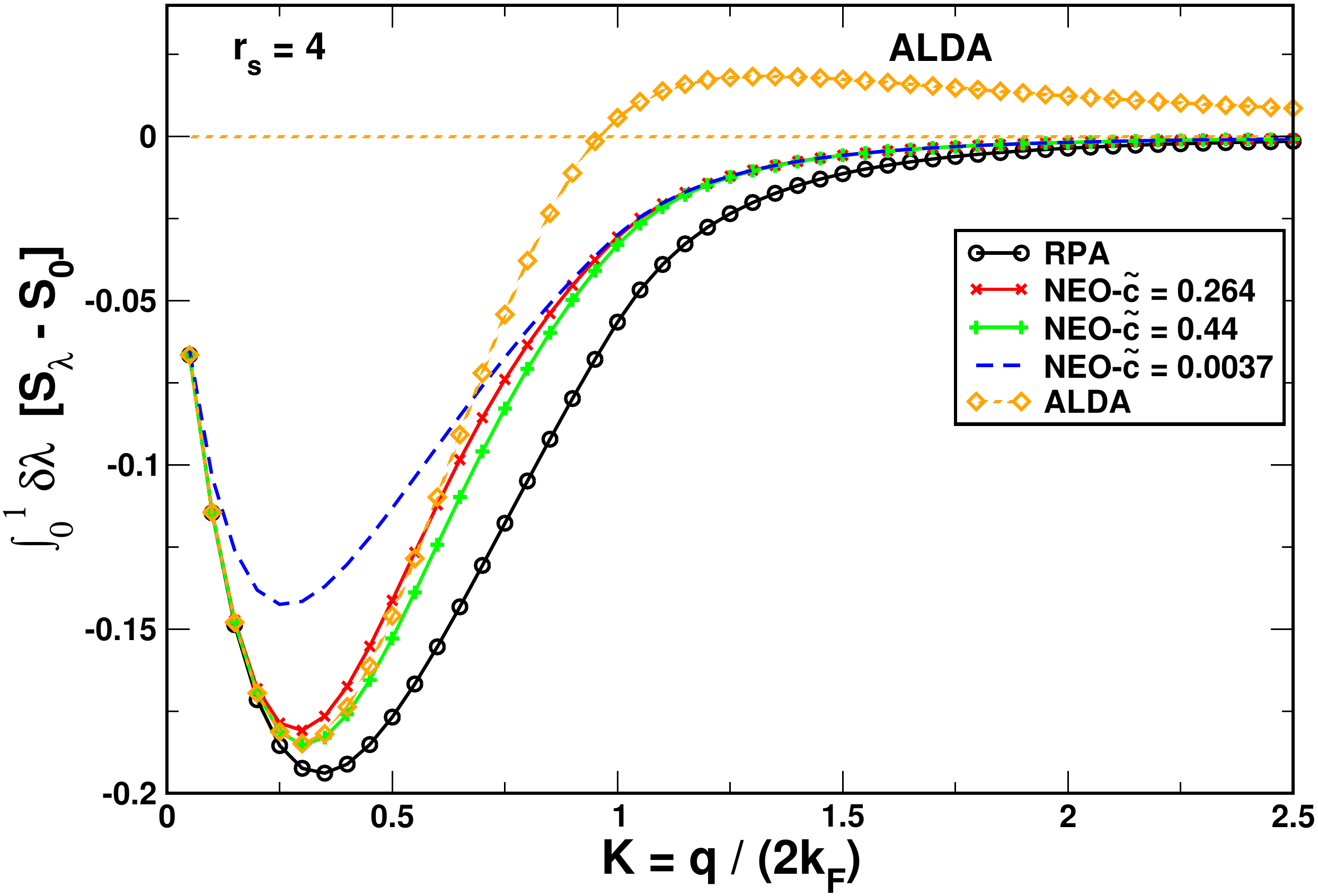}
	\includegraphics[scale=0.3]{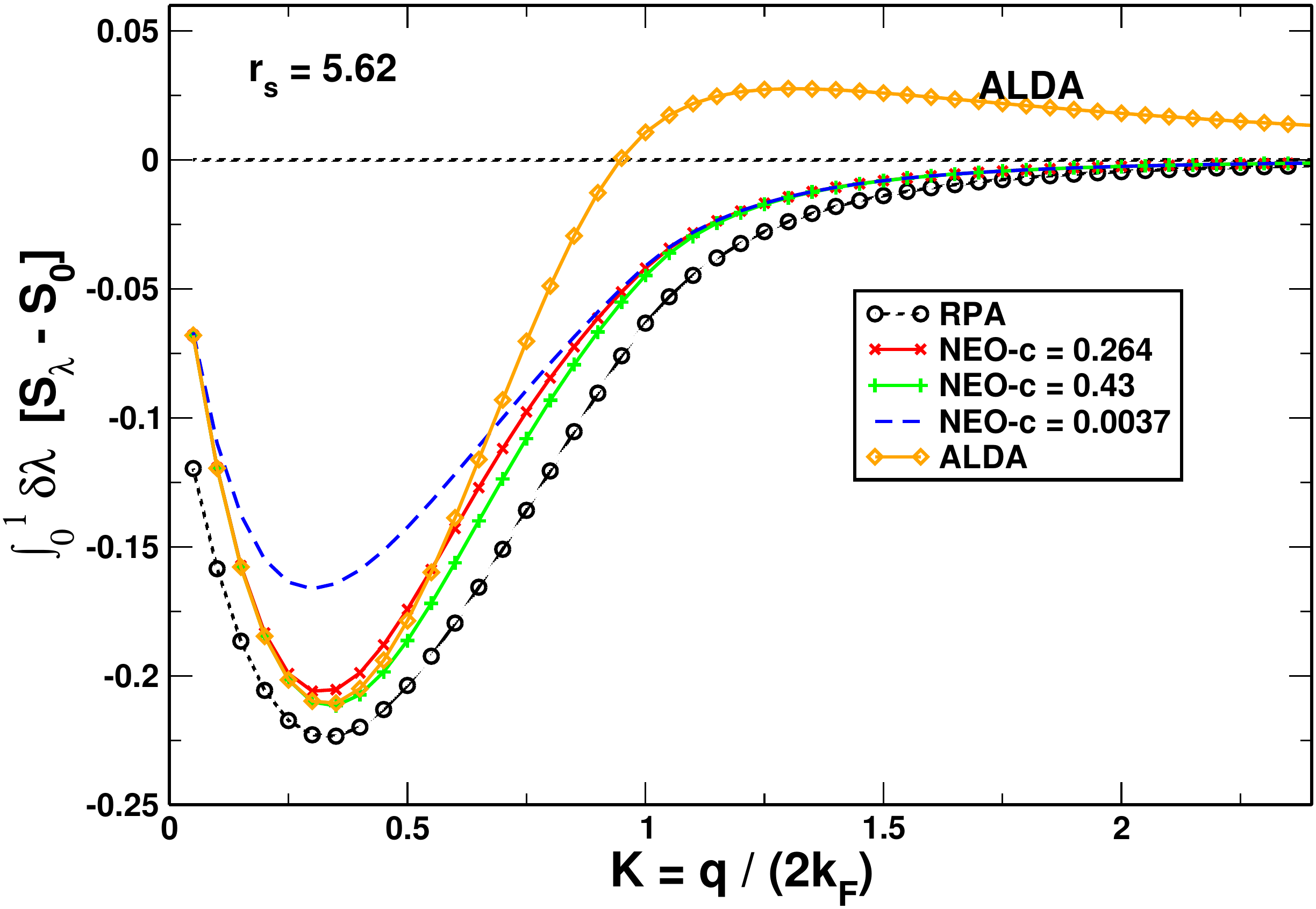}
	\caption{Wavevector analysis of the ground state correlation-only energy of jellium from the dynamic structure factor for reduced wavevector  \( K=\frac{q}{2k_{F}}. \) The area under each curve is proportional to the correlation energy. The left figure shows the correlation-only energy for RPA, ALDA and NEO with the three choices for  \( \widetilde{c}, \)  for r\textsubscript{s}=4. The right figure shows the same for r\textsubscript{s}=5.62 corresponding to Cs.}
	\label{fig7}
\end{figure}

Exchange-correlation kernels can be applied to improve the ground state correlation energy of RPA through the adiabatic connection fluctuation dissipation theorem. This is the basis of the wavevector decomposition of the ground state exchange-correlation energy as known from Langreth and Perdew \cite{LP77}:

\begin{equation}
 E_{xc}= \int _{}^{}\frac{d^{3}q}{ ( 2 \pi ) ^{3}}\frac{1}{2} \int _{0}^{1}\frac{d \lambda }{ \lambda } ( \frac{4 \pi  \lambda }{q^{2}} ) N [ S_{ \lambda } ( {q} ) -1 ],
 \label{eq10}
\end{equation}

\noindent where $ \lambda $  is the coupling constant along the adiabatic connection path and  \( S_{ \lambda } \left( {q} \right)  \)  is the static structure factor found by integrating S$_\lambda$ (q, $\omega$) over frequency and dividing by the electron number. According to the expression given by Eq.~\ref{eq10}, the exchange-correlation energy depends upon the dynamic structure factor. The exchange energy E$_x$ replaces S$_\lambda$ by S$_0$, and the correlation energy is E$_c$ $=$ E$_{xc}$ - E$_x$. Figure~\ref{fig7} shows the wavevector decomposition of the correlation energy for all exchange and exchange-correlation kernels considered here. We plot this for r\textsubscript{s} = 4 and r\textsubscript{s} = 5.62.\\ 

The physical basis of our analysis is the exact exchange-correlation kernel   \( f_{xc} \left( q, \omega  \right)  \)  of the uniform electron gas. For the correlation energy, the static version of the kernel  \( f_{xc} \left( q,0 \right)  \)  can be applied to a good approximation \cite{LGP00}. Note that the frequency dependence at least qualitatively can also be ignored for \(  \omega  \approx  \omega _{p} \)  \cite{IG87, GK85}. The ALDA exchange-correlation kernel  \( f^{ALDA}_{xc} \left( q,0 \right)  \) approaches the exact kernel for the uniform electron gas at  \( q \rightarrow 0 \). In the long wavelength limit,

\begin{equation}
\lim\limits_{q \rightarrow 0} f_{xc} ( q,0) = \lim\limits_{q \rightarrow 0}f_{xc}^{ALDA} ( q,0 ).
 \label{eq11}
\end{equation}

\noindent As we will see below, f$^{ALDA}_{xc}$ breaks down for q/k$_F$ $\geq$ 1, where our constraint-based NEO kernels become less negative and more accurate. Therefore, as suggested by Fig.~\ref{fig0}:

\begin{equation}
f_{xc}^{ALDA} ( q,0 ) <f_{xc} ( q,0 ) < 0
\label{eq12}
\end{equation}

\noindent The ALDA approximation becomes a lower bound to the static exact exchange-correlation kernel for the correlation energy of the uniform electron gas. Figure ~\ref{fig7} visualizes the relation between ALDAxc, RPA and some other exchange-correlation kernels.  The ALDAxc is shown in the left panel of Figure ~\ref{fig7}. The ALDAxc is very accurate for small wavevectors but starts to deviate from NEO at q $=$ k$_F$. At \(  q=2k_{F} \)  the ALDAxc yields a strong overestimation of the correction to RPA correlation energy. All static beyond-RPA kernels make the exchange-correlation energy of RPA less negative for any density including r\textsubscript{s} = 5.62. The correlation energies from NEO  \( \widetilde{c}=0.264 \)  and from the NEO kernel fitted against the compressibility sum-rule  are close to each other but the unphysical NEO  \( \widetilde{c}=0.0037 \)  adds a much larger correction to the RPA correlation energy.\\

The constraints of Equations~\ref{eq11} and~\ref{eq12} control the plasmon dispersion and the correlation energy. The dynamic structure factor becomes a link that couples the physics in the correlation energy and plasmon dispersion. From the correlation energy, the exact uniform-electron gas-based kernel must be less negative than the ALDAxc kernel. The NEO  \( \widetilde{c}=0.264 \)  kernel is uniform electron gas-based only through its energy optimization to the high-density limit, and performs reasonably for both plasmon dispersion and correlation energy in jellium.


\section{Conclusion}
We have presented various model exchange-correlation kernels beyond-RPA for the plasmon dispersion within the jellium model for alkali metals. We have shown that the plasmon dispersion is strictly controlled by exact constraints. Additional physics beyond the ALDA kernel, such as nonlocality in space, can be unimportant for plasmon dispersion. In change, physical constraints such as the compressibility sum rule determine the plasmon dispersion with exchange-correlation kernels. Clearly none of our methods based on particle-hole RPA for jellium is able to predict the experimentally observed negative plasmon dispersion for the heavy alkali metal Cs which arises from band structure. The current exchange-correlation kernels do not have an explicit density dependence that could have a larger impact. For the exact exchange-correlation kernel, the ALDAxc is likely a lower bound (as suggested by Fig.~\ref{fig0}). The ALDAxc is accurate for  \( \frac{q}{k_{F}}<1, \)  the range of  \( q \)  that determines the plasmon dispersion, even though the ALDAxc kernel fails badly for  \( \frac{q}{k_{F}} \gg 1 \), a range that contributes significantly to the correlation energy.

\section{Acknowledgment}
 
The research of N. K. N., S. R. and A. R. was supported by the National Science Foundation under Grant {\fontsize{10pt}{12.0pt}\selectfont No.DMR-1553022}. We thank John P. Perdew for comments on the manuscript.

\noindent$^*$ niraj.nepal@temple.edu\\
$^\dagger$ Corresponding author: aruzsinszky@temple.edu\\

\end{document}